\documentclass[number,sort&compress]{elsarticle}

\usepackage{amssymb}
\usepackage{cancel}
\usepackage{amsthm}
\usepackage{fancyhdr}
\usepackage{lastpage}
\usepackage{caption}
\usepackage{booktabs}

\usepackage{lineno}

\usepackage[figuresright]{rotating}
\usepackage{cancel}

\def \met {\mbox{${\not}{E_T}$}}
\def \metr {\mbox{${\not}{E_T^{raw}}$}}

\def\fb1{~fb$^{-1}$}

\def\gc{~GeV$/c$}
\def\gc2{~GeV$/c^{2}$}

\begin{document}

\begin{frontmatter}



\title{Support Vector Machine Classification on a Biased Training Set: Multi-Jet Background Rejection at Hadron Colliders}
\author{Federico Sforza$^{1,}$\corref{cor1}}
\author{Vittorio Lippi$^2$}
\address{$^{1}$Max-Planck-Institute f\"ur Physik, M\"unchen, Germany}
\address{$^{2}$Uniklinik Freiburg, Freiburg, Germany}
\cortext[cor1]{Corresponding author e-mail: federico.sforza@cern.ch}

\begin{abstract}
  This paper describes an innovative way to optimize a multivariate classifier, in particular a Support Vector Machine algorithm, on a problem characterized by a biased training sample. 
  This is possible thanks to the feedback of a signal-background template fit performed on a validation sample and included both in the optimization process and in the input variable selection. The procedure is applied to a real case of interest at hadron collider experiments: the reduction and the estimate of the multi-jet background in the $W\to e \nu$ plus jets data sample collected by the CDF experiment. 
The training samples, partially derived from data and partially from simulation, are described in detail together with the input variables exploited for the classification. At present, the reached performance is superior to any other prescription applied to the same final state at hadron collider experiments.

\end{abstract}

\end{frontmatter}

\thispagestyle{myheadings}

\section{Introduction} 
\label{Intro}

A multivariate classifier is an adaptive algorithm trained to identify a signal of interest against other background events on the basis of a set of input variables. Therefore the understanding of the training samples and the input variables selection are two key elements to obtain optimal results.

In this paper we apply the previous paradigm in an innovative way to both the training and the input variable selection of a Support Vector Machine~\cite{BShop_machine_learning} (SVM) algorithm. In particular we obtain an excellent signal-background multivariate classifier when one of the training samples is biased  (i.e. it does not correctly reproduce all characteristics of the signal or of the background samples) and statistically limited to few thousands of events.

We decided to explore the use of the SVM algorithm (described in Section~\ref{sec:SVM}) because of several advantages with respect to other multivariate techniques. For example, Artificial Neural Networks, commonly used in High Energy Physics~\cite{PhysRevD.54.1233}, require to arbitrarily set the complexity of the classifier (i.e. the number of neurons and layers of the net), the training may converge to local minima and, usually, large training sets are needed to finely map the input space. On the other hand the SVM algorithm, whose basic idea is the identification of the best hyper-plane separating two classes of vectors, has unique solution of the training algorithm, a small number of free parameters and good performance on low statistics training sets, as only a small number of training vectors are exploited in the final solution~\cite{svmbook}. Other promising results using SVMs in High Energy Physics analyses are reported in Ref.~\cite{A.2003, PC2003327, sanctis:105}. 

In this work, developed in the framework of the analysis searching for the Higgs boson with the Collider Detector at Fermilab (CDF) experiment~\cite{wh75_PRD}, we deal with three machine-learning challenges: 
the reduction of the effect of a bias in the training set,
a robust evaluation of the performance of the trained SVM, and the optimal input variable selection. 

The key point to achieve all of the above is the scan of the free parameters of the training procedure together with a cross check of the efficiency of each resulting SVM (described in  Section~\ref{sec:training_set}). The efficiency cross check is performed both on the training set, with a $n$-fold cross validation, and on unclassified events, with a template fitting procedure over the SVM output value distribution (described in Section~\ref{sec:method}).

We tested the developed methodology on a toy model (Section~\ref{sec:toy}) and finally we applied it to a real physics case (Section~\ref{sec:cdf_set}). In this latest part we exploited the described template fitting procedure to identify a reliable and optimal input variable selection. A wide literature discusses the topic of input variable selection but we focused on the identification of the best, minimal set of inputs. It is clear that a highly discriminative variable will improve the classification power but it has been shown in Ref.~\cite{Weston00featureselection} that also the performance of the SVM algorithm itself improves when the variables are well chosen, especially if they have very different discrimination power~\cite{Chen05combiningsvms}. Intuitively, the introduction of too many not-significant inputs introduces a noise term to the algorithm, decreasing its performance. Furthermore, a reduced set of variables identifies the most important characteristics of the analyzed process and may decrease the number of needed validation checks. The complete procedure (described in Section~\ref{sec:var_sel}) involves the automatic training and performance evaluation with several variables configurations.

The physics case under exam is the reduction and estimate of the multi-jet background contamination in a dataset enriched in leptonic $W$ bosons decays, selected in association with hadronic jets. 

The dataset, collected at a proton-antiproton collider which was operating at $\sqrt{s}=1.96$~TeV (the Tevatron), is one of the main investigation channels at hadron collider experiments. Several interesting but rare processes (i.e. associate $WH$ production, diboson or top quark production, etc.) produce a $W$ boson in the final state. The $W$ leptonic decay is used to identify a clear event signature out of the overwhelming multi-jet background produced by generic QCD interactions between the proton and antiproton constituents. Hadronic jets faking the lepton and the neutrino identification introduce a significant multi-jet background contamination (especially for electrons identification algorithms). Because of its nature, the multi-jet background is a mixture of detector effects and physics processes. Usually data-driven models obtained with a specific selection enriched in multi-jet events, are used to estimate this contamination. These models may be statistically limited and the use of a different selection criteria often introduces unexpected biases in the simulated variables. All this makes the application of multivariate techniques particularly challenging.

\section{Support Vector Machines}\label{sec:SVM}

The SVM is a supervised learning binary classifier whose basic function is the identification of the {\em best separating hyper-plane} between two classes of $n$-dimension vectors. 

Given a training set made by two classes of vectors (i.e. signal and background) linearly separable, the SVM algorithm produces, as a solution, a unique plane defined by the vectors at the boundary of the two classes; those are the so called {\em support vectors}. In the case of non-linear separation, the plane is found in an abstract space, defined by a transformation of the input vectors. Although the transformation can be very complex, it is not necessary to know it exactly, but we just need to know its effect on the scalar product between the vectors, named {\em Kernel}. Finally the cases of not perfect separability of the two samples are solved by introducing a penalty parameter accounting for the contamination. 

It is possible to find more details in Ref.~\cite{svmbook} and~\cite{BShop_machine_learning}, but, for sake of clarity, a short overview of the algorithm is also given in the following. For the actual, numerical, implementation of the SVM algorithm we relied on the \texttt{LIBSVM} open source library~\cite{libsvm}.

\subsection{The Linear Case}
\begin{figure}
  \begin{center}  
    \includegraphics[width=1.0\textwidth]{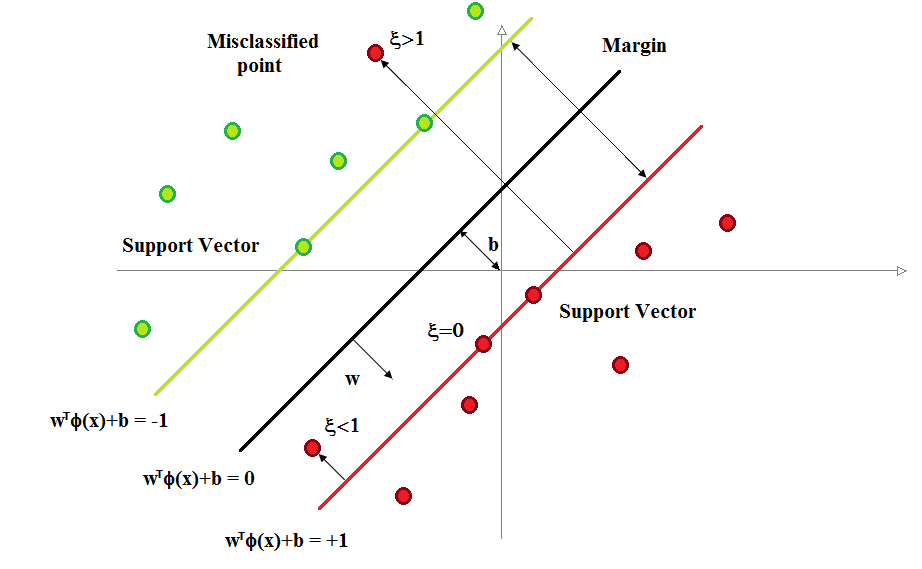}
    \caption[Example of SVM Linear Separation]{An example of SVM: two linearly separable classes of vectors are represented with red and blue dots. The plane leading to a maximum separation is defined by the weight vector $\vec{w}$ and the constant term $b$.}\label{fig:svm_linear}
  \end{center}
\end{figure}

Figure~\ref{fig:svm_linear} shows how the linear classification problem can be formalized in the  minimization of $|\vec{w}|^2$ (with $\vec{w}$ = vector normal to a plane) with the constraint:
\begin{equation}\label{eq:constraint}
  y_i(\vec{x_i}\cdot\vec{w} + b)-1\geq 0\qquad
  \left\{\begin{array}{ll}
  y_i=+1, & i \in \textrm{signal\,;} \\
  y_i=-1, & i \in \textrm{background\,;}\\
  \end{array} \right.
\end{equation}
the problem has a unique solution obtained by the maximization of:
\begin{equation}\label{eq:lagrange_1}
  L=\sum_i \alpha_i - \frac{1}{2}\sum_{i,j}\alpha_i\alpha_j y_i y_j\vec{x_i}\cdot\vec{x_j}\,,
\end{equation}
which are derived by the application of the Lagrange multipliers to Eq.~\ref{eq:constraint}.

The solution identifies $\alpha_i>0$ for some $i$. The associated vectors, i.e. the support vectors, are a subset of the training sample that defines the best hyper-plane separating the two input classes (see Figure~\ref{fig:svm_linear}).

When not completely separable classes of vectors are present, a {\em penalty parameter} $C$ is added to account for the contamination. The new minimization condition is:
\begin{equation}
  |\vec{w}|^{2} + C \sum_{i}{\xi_{i}}\,;
\end {equation}
with the new constraint (derived from Eq.~\ref{eq:constraint}):
\begin{equation}\label{eq:lagrange_2}
  y_i(\vec{x_i}\cdot\vec{w} + b) \geq 1-\xi_{i}\quad\textrm{with}\quad \xi \geq 0\textrm{\,.}
\end{equation}
The parameter $C$ defines the SVM implementation before the training, therefore it represents a hyper parameter of the SVM.

For any new vector $\vec{X}$ considered for classification, we evaluate its position $D(\vec{X})$ with respect to the plane defined by the support vectors $\vec{x_i}$ and the parameters $\alpha_i$:
\begin{equation}\label{eq:distance}
  D(\vec{X})= \sum_i\alpha_i y_i \vec{x_i}\cdot \vec{X} - b\,,
\end{equation}
where $b$ is a constant term of the solution. 

The variable $D$ is the final output of the SVM: its sign gives the signal-background classification and, if the geometry is simple, its value is the distance of a test vector $\vec{X}$ from the classification plane. As we are going to see in the next paragraph, a non-linear classification is possible only thanks to a not-explicit transformation to a different vector space, where $D$ may loose the immediate geometrical meaning. 

Natively SVMs are used as binary classifiers but, here, we add a large degree of flexibility by exploiting the full information of the continuous variable $D$. The SVM is used as a dimensionality reducer of the classification problem: the separation power and the correlations among several significant input variables are summarized into one continuous distribution.

\subsection{Kernel Methods}

Non-linearly-separable classes of vectors can be classified by transforming them with an appropriate function, $\Phi(\vec{x})$, that maps the elements into another space, usually with higher dimension, where the linear separation is possible. 

However the  identification of $\Phi(\vec{x})$ is not trivial, therefore the so called {\em Kernel trick} is often used. A Kernel function, $\mathbf{K}(x_i,x_j)$, generalizes the scalar product appearing in Eq.~\ref{eq:lagrange_1} (or Eq.~\ref{eq:lagrange_2}) without the need of explicitly knowing $\Phi(\vec{x})$. The Kernel is the composition of the mapping $\Phi(\vec{x})$ with the inner product:
\begin{equation}
  \mathbf{K}(x_i,x_j)= \Phi(x_i)\cdot\Phi(x_j)\quad\mathrm{with}\quad \Phi : \Re^n\mapsto \mathcal{H}\,.
\end{equation}
The function $\mathbf{K}$ should satisfy to a general set of rules to be a Kernel, but we describe only the {\em Gaussian} Kernel we used in this work. It is expressed as:
\begin{equation}\label{eq:gauss_k}
K(x_{i},x_{j}) = e^{-\gamma |\vec{x}_{i}-\vec{x}_{j}|^2}\,;
\end{equation}
The corresponding $\Phi(x)$ is unknown and it maps the input vectors to an infinite dimension space. The Kernel is defined only by one hyper-parameter, $\gamma$, that should be set before the training of the SVM.

\section{SVM Training on a Biased Sample}\label{sec:method} 

Several multivariate techniques are based on the assumption that the labelled samples used for the classifier training are drawn from the same probability distribution of the unclassified events. 
In our case of study, where only an approximate and statistically limited model of the background processes is available (see Section~\ref{sec:cdf_set} and in particular Section~\ref{sec:training_set} for the multi-jet background description), we do not expect the previous assumption to hold for every portion of the phase space. To cope with this problem, we developed an original methodology to evaluate the SVM training performance.

Section~\ref{sec:SVM} shows that, for each choice of hyper-parameters and training vectors, only one optimal SVM solution exists and we need to evaluate the performance of it. 

As a performance estimator we use the {\em confusion matrix} of the classifier: the element $(i,j)$ of the matrix is the fraction of the class $i$ classified as member of class $j$. Figure~\ref{fig:confusion} shows a representation of it in the two-classes case, where one class is labelled as {\em background} and the other as {\em signal}. We obtain a reliable estimate of the classifier quality by filling the confusion matrix in two independent ways and combining all the available information.

\begin{figure}[h!]
  \begin{center}   
{
\renewcommand{\arraystretch}{1.8}
    \begin{tabular}{|c|c|}
      \hline
      {\em Sgn} classified as {\em Sgn} & {\em Bkg} classified as {\em Sgn}\\
      \hline
      {\em Sgn} classified as {\em Bkg} & {\em Bkg} classified as {\em Bkg}\\
      \hline
    \end{tabular}   
}
    \caption[Definition of Confusion Matrix]{Definition of confusion matrix for a two classes ($Sgn$ and $Bkg$) classification problem. This reproduces the case of an algorithm used to discriminate signal {\em vs} background: the elements of the matrix are the signal and background classification performance and the cross contamination.}\label{fig:confusion}
  \end{center}
\end{figure}

The first performance evaluation method is the {\em $k$-fold cross-validation}: the training set is divided into $k$ sub-samples of which one is used as a validation set and the remaining $k - 1$ are used in the training; the confusion matrix is then evaluated applying the trained discriminant to the validation set. The cross-validation process is repeated $k$ times (the {\em folds}) and the final performance is the average of them. This method is solid against over fitting but it has no protection against biases on the training sample.

The second method, a key feature of this work, exploits a signal plus background template fit to extract the off-diagonal terms of the confusion matrix. 
The fit is performed on a validation sample of unclassified events (i.e. the {\em data}) using a significant variable that allows some signal to background discrimination. While the signal and background templates are derived directly from the training samples, the unclassified data events are composed by an unknown mixture of true signal and background events. 

The fitting routine, implemented in the ROOT~\cite{rootr} analysis package and derived from Ref.~\cite{Barlow_1993}, maximizes a binned likelihood function, $\lambda$, over the significant variable distribution. The fractions of the signal and of the background templates are the free parameters from which we derive the elements of the confusion matrix.

If the variable chosen for the fit is not well reproduced by the simulation then we expect that the fitted fractions are going to largely differ from the results obtained with the $k$-fold cross-validation. At the same time we can quantitatively evaluate the agreement between the data shape and the fitted templates as the quantity:
\begin{equation}
  \chi^2 = -2\ln(\lambda)
\end{equation}
follows a $\chi^2$ probability distribution (under general assumptions).

The last critical point is the identification of a sensitive variable to be used in the fit. In a previous work~\cite{CHEP_svm} we used the distribution of the imbalance of the total transverse momentum of the particles produced in a hadron collision (also referred as missing transverse energy or \met), as it is sensitive to the background we were interested in (i.e. multi-jet contamination). Here we moved to a more general approach, by the machine learning point of view, with the direct usage of the SVM output value $D$, defined by Eq.~\ref{eq:distance}. We suppose that, if the SVM training performance evaluation is reliable, the variable $D$ is highly sensitive to the background and signal composition of the data sample, therefore it can be used in the template fit procedure. To verify the reliability of the training, we evaluate the $\chi^2$ of the template fit: this ensures a good shape agreement between the data and signal and background templates. We also cross checked the validity of the fit procedure over a toy example discussed in the following.

\begin{figure}[!h]
  \begin{center}
  \includegraphics[width=0.9\textwidth]{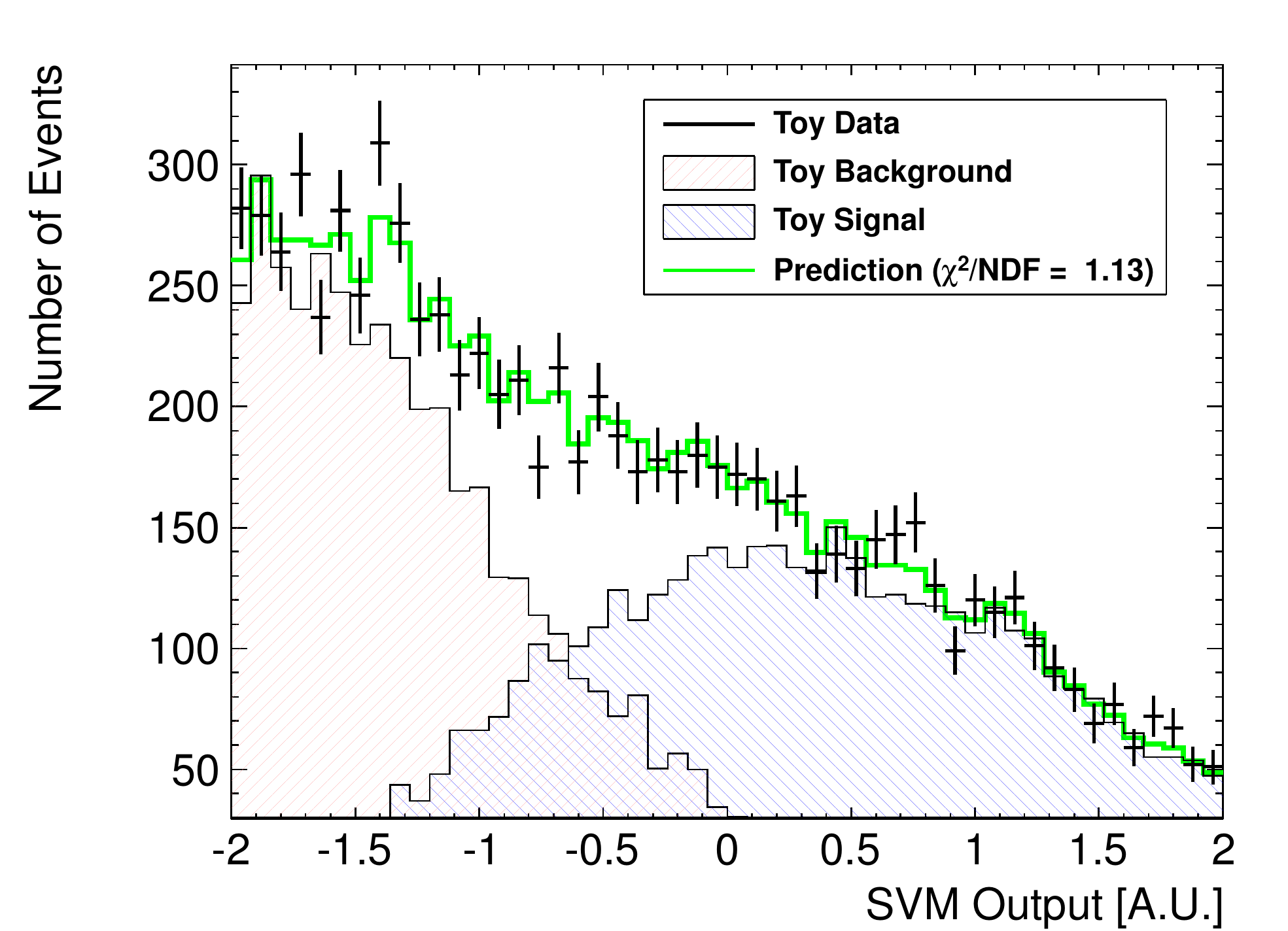}
  \caption[Two-component Template Fit on Toy Data]{A two-component fit of signal (blue) and background (red) templates is performed for the distribution of the SVM variable $D$ (Eq.~\ref{eq:distance}), over toy generated data.}\label{fig:toysamplefit}    
  \end{center}
\end{figure}

\subsection{A Toy Example}\label{sec:toy} 

We built a toy example in order to verify the robustness of the proposed method for an SVM performance evaluation when partially biased samples are present. The toy is composed by three data-sets generated with known probability distributions:
\begin{description}
\item[signal model:] $10^5$ vectors generated from a $2-$dim Gaussian distribution with the following mean, $\vec{\mu}_{Sgn}$, and standard deviation, $\tilde{\sigma}_{Sgn}$:
  \begin{equation}\label{eq:sgn_gaus}
    \vec{\mu}_{Sgn} = \left[
      \begin{array}{c}
        - 3  \\
        0  \end{array} \right]
    ,\qquad 
     \tilde{\sigma}_{Sgn} = \left[
       \begin{array}{cc}
         8 & 0 \\
       0 & 8  \end{array} \right].
  \end{equation}
  
 \item[Background model:] $10^5$ vectors generated from a $2-$dim Gaussian distribution with the following mean, $\vec{\mu}_{Bkg}$, and standard deviation, $\tilde{\sigma}_{Bkg}$:
   
   \begin{equation}\label{eq:bkg_gaus}
     \vec{\mu}_{Bkg} = \left[ 
     \begin{array}{c}
       3  \\
       0  \end{array} \right]
 ,\qquad 
     \tilde{\sigma}_{Bkg} = \left[
     \begin{array}{cc}
       8 & 0 \\
       0 & 8  \end{array} \right].
   \end{equation}
 \item[Data:] a mixture of $5 \cdot 10^4$ vectors generated from the same distribution of the signal model (Eq.~\ref{eq:sgn_gaus}) and $5\cdot 10^4$ vectors generated from a background distribution similar to the background model (Eq.~\ref{eq:bkg_gaus}) but with $\hat{\sigma}_{Bkg}$ increased by 20\% in one direction to simulate a mismatch between the real background and the model.
 \end{description}
We tested several combinations of the hyper-parameters $C$ and $\gamma$ using the signal and background model in the training. For each obtained SVM we evaluated the $k$-fold cross validation and we performed the template fit on the SVM variable $D$ evaluated on the data sample. Figure~\ref{fig:toysamplefit} shows an example of the fit.

\begin{figure}[!h]
   \begin{center}
   \includegraphics[width=1.0\textwidth]{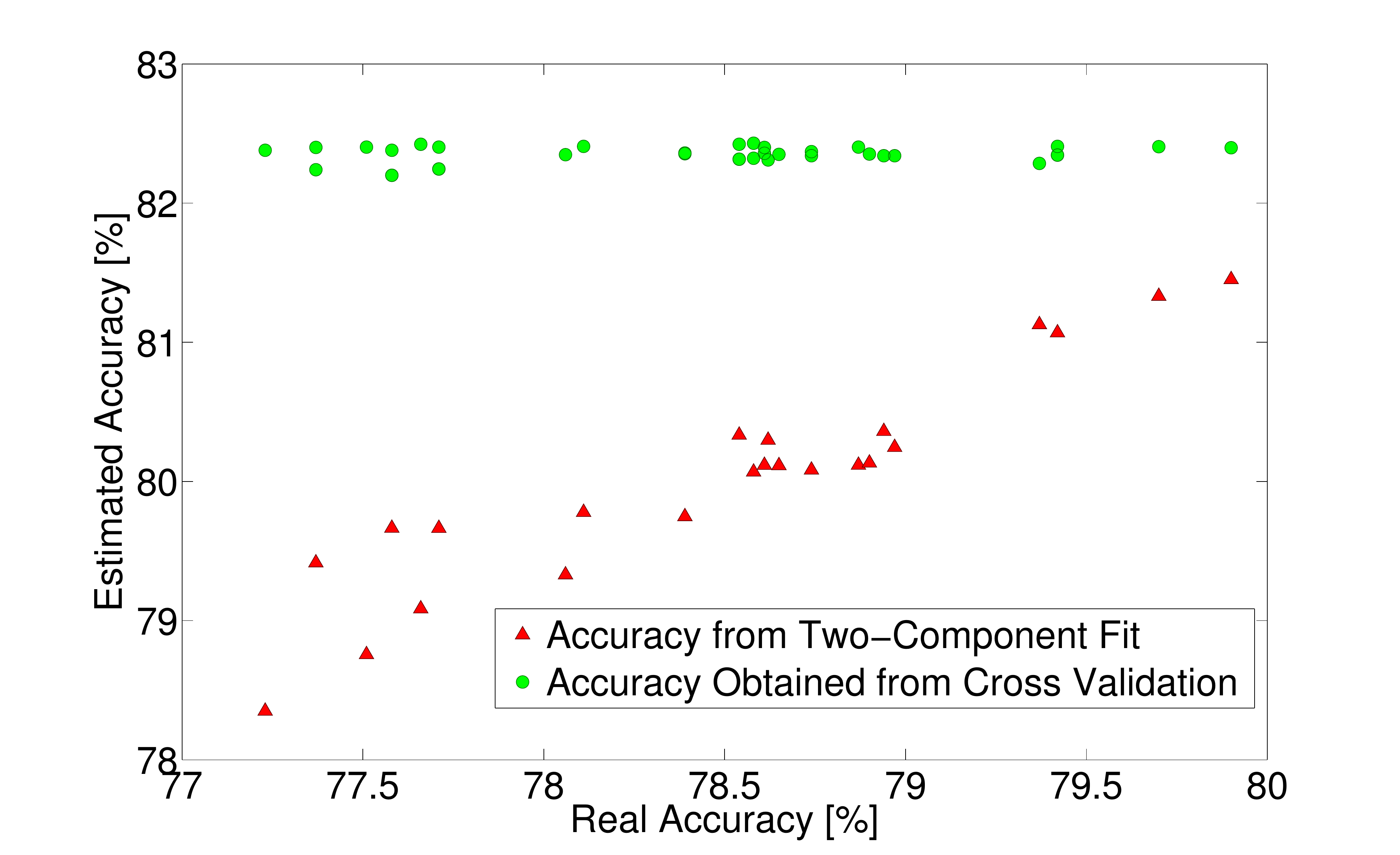}
   \caption[Toy Model SVM Performance Estimate with Fit and Cross-Validation]{SVM performance estimate with a $k$-fold cross-validation (green circles) and with a two-component signal and background template fit on the SVM output variable (red triangle) of toy data of known composition. The true performances of the SVM classifier are reported on the $x$ axis. The fit evaluation appears on the diagonal of the plane, signaling a realistic estimate of the true performance of the classifier with respect to the direct $k$-fold evaluation which biases the result by a sizable percentage.}\label{fig:ToyResult}     
   \end{center}
 \end{figure}

The result is reported in Figure~\ref{fig:ToyResult} and, as we know the true label of the toy data vectors, the real performances of the SVM are reported on the $x$ axis of the diagram. The evaluation of the performance obtained from the fit lies on the diagonal of the plane. It gives a much more realistic estimate of the SVM classifier performance with respect to the direct $k$-fold evaluation which may bias the result by a sizable amount also in this simple case.

\section{Multi-jet Background Rejection in the $W$ plus Jets Data Sample}\label{sec:cdf_set}

The algorithm described in previous sections can be applied to the reduction of the multi-jet background in the $W$ plus jets channel at hadron collider experiments.

This channel is the basis of many relevant analyses. Events are selected by the identification of the leptonic decay of a $W$ boson ($W\to e \nu$ or $W\to \mu \nu$ in our case) together with one or more jets (i.e. final states of a quark hadronization). Several interesting processes are detectable in such final state but they are characterized by a tiny production cross section ($\mathcal{O}(1)$~pb) if compared to the total $p\bar{p}$ inelastic collision cross section ($\mathcal{O}(1)$~mb) at $\sqrt{s}=1.96$~TeV. A few examples: Higgs boson production in association with a $W$ boson ($WH$), single-top production, diboson production ($WW$, $WZ$). The selection of a leptonic $W$ boson decay is the key of the signal identification as it reduces the rate of uninteresting processes by a factor of $10^5$ with respect to the total inelastic cross section of $\approx 10^8$~pb, typical of the $p\bar{p}$ interactions at the, $\sqrt{s}=1.96$~TeV, Tevatron center of mass energy. 

A mixture of physics processes and detector effects may produce the reconstruction (mis-identification) of a fake $W$ boson thus allowing contamination of the selected sample by multi-jet events.
In general~\cite{single_top}, such background is reduced by a more accurate lepton identification or by rejecting events with kinematic not compatible with a $W$ boson decay. The remaining contamination is then estimated with data-driven models where some of the lepton selection requirements are inverted to obtain a multi-jet enriched sample. The modeling and the understanding of the sample remain a challenge because of two main reasons: first the multi-jet models are statistically limited to the actual selected data, second, the inversion of some selection criteria may bias the sample.
Both these effects pose strong constraints on the applicability of multivariate techniques on the multi-jet rejection.

\begin{figure}[ht]
  \begin{center}
    \includegraphics[width=1.0\columnwidth]{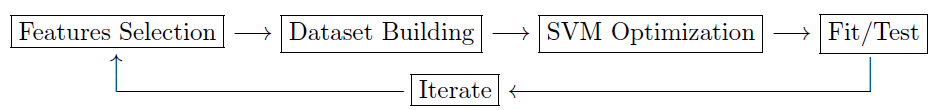}
    \caption{Flowchart of feature selection - training - test procedure.}\label{flowchart}
  \end{center} 
\end{figure}

Our SVM classifier overcomes these difficulties because the optimization and the training take into account a cross check on an unclassified data control sample. The preliminary idea of this algorithm~\cite{CHEP_svm, phx2011} proved to be successful in several analyses performed by the CDF collaboration~\cite{wh75_PRD, wh94_PRL, cdfDiblvHF_Mjj2011, cdf_zprime2012}. In previous works the SVM was used, following its original concept, as a binary classifier. In this paper we discuss a more powerful and innovative use of the algorithm. We use the continuous distribution $D$ (Eq.~\ref{eq:distance}) explicitly in the optimization. The improvement is dual: although the training algorithm sets the optimal signal selection above $D = 0$, now the threshold level can be varied to increase the signal efficiency or decrease the background contamination, depending on the physics analysis needs. Furthermore the agreement of the SVM output distribution with data can be used to extract information about the multi-jet modeling and normalization. We heavily relied on this second feature in the variable selection process. As described in section~\ref{sec:var_sel} and schematized in Figure~\ref{flowchart}, we trained a new, optimal SVM for each variable configuration, then we tested the performance and the agreement of the SVM using the $D$ variable and the two-component fit described in Section~\ref{sec:method}. The process was automatically iterated until stable performance was achieved.

To prove the robustness and the quality of the complete algorithm, we applied it to two datasets, collected by the CDF II experiment and characterized by different kinematic and background contamination. The first contains a high energy electron identified in the central region of the detector and the second contains a high energy electron identified in the forward region of the detector. We chose to train the algorithm only on the electron sample because the multi-jet contamination is expected to be larger than in the muon sample. However, as we do not use specific lepton identification variables but we base the discrimination on the event kinematic, the same algorithm proved to be optimal also for several other lepton categories~\cite{fede_sforza_tesi}.

\subsection{Training Sample Description and Selection Criteria}\label{sec:training_set}

We built the training sets using electron plus jets events selected in data and Monte Carlo (MC) samples with the standard particle identification algorithms used by the CDF Collaboration~\cite{top_lj2005}.

A brief description of the experimental setup is helpful to understand the selection criteria. The CDF II is a general-purpose particle detector placed at one of the two collision points of the Tevatron $p\bar{p}$, $\sqrt{s}=1.96$ collider (in operation from 2001 to September 2011).
Different information on the data is collected by several subdetectors: a tracking system (silicon detector~\cite{svx} plus drift chamber~\cite{cot} in a $1.4$~Tesla solenoid), a calorimeter system~\cite{cen_cal} (composed by an electromagnetic and a hadronic section) and an outer muon identification system (composed by drift chambers and scintillators~\cite{muon}). The subdetectors have azimuthal and forward-backward symmetry with respect to the geometrical center of the detector corresponding to the nominal collision point. Positions and angles are expressed in a cylindrical coordinate system, with the $z$ axis along the proton beam, azimuthal angle $\phi$ and polar angle $\theta$. The following variables are defined according to these principles\footnote{They are relativistic invariant in the case of massless particles.}: the pseudorapidity \mbox{$\eta = - ln[\tan (\theta/2)]$}, the transverse energy \mbox{$E_T=E \sin\theta$} (as measured by the calorimetry), the transverse momentum $p_T= p \sin \theta$ (as measured by the tracking systems) and the angular distance between two particles in the $\eta - \phi$ space, $R =\sqrt{ (\Delta\eta)^2 + (\Delta\phi)^2}$.

The identification of central electron candidates ($|\eta|<1.1$) is based on the following criteria~\cite{Abulencia:2005ix}: a good quality track pointing to a significant energy deposit in the electromagnetic section of the central calorimeter ($E_T > 20$~GeV) and the compatibility of the electromagnetic shower shape and composition (according to five variables) with test-beam data and $Z\to e\bar{e}$ events. 

Forward electron candidates (identified in the calorimeter region with $1.2<|\eta|<2.0$) are identified in a similar way but, due to poor tracking chamber coverage, no track matching is required and a different strategy, named Phoenix matching scheme~\cite{phx_alg}, is used to reject fake-electrons.

The $W\to e\nu$ identification is completed by requiring the electrons to be isolated from nearby activity, as measured in the electromagnetic calorimeter, within a cone of $R=0.4$ ($Iso = E_{R=0.4}/E_{R=0.1} < 0.1$) and the presence of an imbalance in the total transverse energy measured by the calorimeter system greater than $15~$GeV (\met$>15$~GeV). 
The isolation requirement derives from the expected kinematic behaviour of the $W$ decay while the \met ~signals the presence of a neutrino in the event\footnote{The total transverse momentum of the particles produced in a hadron collider can be considered exactly zero in the center-of-mass system, therefore an imbalance signals the presence of at least one escaping undetected particle as a neutrino.}. Events with another lepton candidate are rejected as they introduce $Z\to e\bar{e}$ contamination.

The final step of the lepton plus jets selection is the reconstruction of two or more central ($|\eta|<2.0$) jets using a fixed cone ($R=0.4$) identification algorithm~\cite{jet_corr1}. 
The transverse energy of the jets should be greater than $18$~GeV, after correcting for detector effects. The per-jet correction, named $Cor_j$ in the following, is also taken into account in the evaluation of \met. Large jet activity and fluctuations in their energy measurement can originate false \met ~thus increasing the probability of $W$ mis-identification. 
In the following we indicate with $raw$ superscript all the quantities calculated before jet energy correction: i.e. \metr ~and \mbox{$E_{T,Jet}^{raw}$}.
An additional requirement of \metr$>20$~GeV is applied in the forward electron sample to remove a bias caused by the online event selection.

The described selection criteria are used to build the training and test samples according to the following specifications:
\begin{description}
  \item[Signal:] both for the central and the forward electron selection, we used a MC-based training set of $W\to e\nu +$jets events. A reliable and simple estimate of the expected signal kinematic and hadronization properties is obtained by combining the $W\to e\nu$ plus one and plus two partons MCs where the events are generated by ALPGEN~\cite{alpgen} and the hadronic showering is performed by PYTHIA~\cite{pythia}. We used for the training only $7000$ events, out of the approximately $10^5$ generated, while we kept the rest for validation purposes. The small size of the signal training samples is forced by the statistical limits of the background samples and by the need to balance them in the training phase.
\item[Background (Central):] the multi-jet model specific to the central electrons selection is obtained by using samples enriched in fake electrons. In particular, the electromagnetic shower shape and composition comparison should fail at least two out of the five requirements.
\item[Background (Forward):] an appropriate multi-jet background training set is obtained, for the forward electron sample, by selecting non-isolated electrons (\mbox{$Iso > 0.1$}).
\item[Unclassified Data:] as explained in Section~\ref{sec:method}, we also need a sample of unclassified events for the cross check of the SVM training performance. To this purpose we used about one fifth of the CDF data ($\mathcal{L}\approx 2$~fb$^{-1}$) corresponding to average run conditions and luminosity profile.
\end{description}
The different prescriptions used for the multi-jet background models were developed in several CDF analyses performed in $W$ plus jets data sample (two relevant examples are in Ref.~\cite{top_lj2005} and~\cite{single_top}) and they are essentially obtained by inverting one or more selection criteria with respect to the standard $W\to e\nu$ selection. 
The total amount of multi-jet background events available was about $1.3\times 10^4$ both for the central and the forward samples.
Before the training, the behaviour of the background models has been studied and improved in few aspects as described in Ref.~\cite{fede_sforza_tesi}.

%

\subsection{Input Variable Description}\label{sec:var_desc}

A multivariate algorithm relies on a given set of input variables. The {\em feature selection} problem is fundamental in machine-learning and, if possible, even more in the present case where the background sample does not guarantee a good model of all the variables. 

We started with a large set of variables (twenty-four) identified according to two basic criteria: no evident correlation with the lepton identification variables and the use of the kinematic difference between the simulated $W+$ jets events and the multi-jet background model. At the end of the optimization, these requirements allowed the use of the multi-jet rejection procedure on several different lepton identification algorithms.

All the variables are listed in Table~\ref{tab:input_vars}. They involve the kinematic properties of the lepton, the leading and second leading jet and the \met ~module and direction. In the following we describe the few more complex variables entering in the set:
\begin{itemize}
\item \mbox{${\not}{p_T}$} is the missing momentum defined as the momentum imbalance on the transverse plane. It is computed adding all the reconstructed charged tracks transverse momenta, $\vec{p_i}$:
  \begin{equation}
    \vec{{\not}{p_T}}\equiv - \sum_i{\vec{p^i}_T}\quad\textrm{with }|\vec{p^i}_T|>0.5\textrm{GeV}/c\textrm{;}
  \end{equation}
\item $M_T^W$ is the {\em transverse mass} of the reconstructed $W$ boson:
  \begin{equation}
    M_T^W=\sqrt{2(E_T^{lep}\cancel{E}_T - E_x^{lep}\cancel{E}_x - E_y^{lep}\cancel{E}_y)}\mathrm{.}
  \end{equation}
\item $MetSig$ is the \met ~{\em significance}, a variable that relates the reconstructed \met ~with the detector activity (jets and unclustered energy):
  \begin{equation}
    MetSig =\frac{\cancel{E}_T}{\sqrt{ \Delta E^{jets}+ \Delta E^{uncl}}}\mathrm{,}
  \end{equation}
  where:
  \begin{eqnarray}
    \Delta E^{jets} =\sum_{j}^{jets}\Big(Cor_{j}^2\cos^2\Big(\Delta\phi\big(\vec{p_j},\cancel{\vec{E}}_T\big)\Big)E_{T}^{raw, j}\Big)\mathrm{,}\\
    \Delta E^{uncl} = \cos^2\Big(\Delta\phi\big(\vec{E}_T^{uncl}\cancel{\vec{E}}_T\big)\Big)E_T^{uncl}\mathrm{,}
  \end{eqnarray}
  $uncl$ refers to the calorimeter energy not clustered into electrons or jets and $Cor_j$ is the total correction applied to each jet.
\item $\nu^{Min},\quad \nu^{Max}$ are the two possible reconstructions of the neutrino momenta. As the $p_z^{\nu}$ component is not directly measurable we infer it from the $W$ boson mass and the lepton momentum. The constraints lead to a quadratic equation which may have two real solutions, one real solution, or two complex solutions\footnote{The real part is chosen in this case.}. The reconstructed $\nu^{Min}$, $\nu^{Max}$ derive from the distinction of $p_z^{\nu,Max}$ and $p_z^{\nu,Min}$.
\end{itemize}

\begin{table}
{
  \begin{tabular}{cl| cl|  cl| cl }
    \toprule
    \multicolumn{8}{c}{Possible Input Variables }\\
    \midrule
    1 & $p_T^{lep}$           &  7 & $E_T^{raw, jet1} $   & 13 &  $\Delta\phi( {\not}{p_T},$ $lep)$ &   19 & $\Delta R( lep,$ $jet 2)$ \\
    2 & \met                 &  8 & $E_T^{raw, jet2} $   & 14 &  $\Delta\phi( {\not}{p_T},$ \met$)$ & 20 & $\Delta R( \nu^{min},$ $jet 1)$ \\ 
    3 & \metr                &  9 & $E_T^{cor,jet1} $       & 15 & $\Delta\phi( {\not}{p_T},$ \metr$)$ &  21 & $\Delta R( \nu^{min},$ $jet 2)$ \\ 
    4 & \mbox{${\not}{p_T}$} & 10 & $E_T^{cor,jet2} $       &  16 & $\Delta\phi( lep,$ \met$)$ &    22 & $\Delta R( \nu^{min},$ $lep)$ \\ 
    5 & $M_T^W$              & 11 & $\Delta\phi( jet 1,$ \met$)$ & 17 & $\Delta\phi( lep,$ \metr$)$ &  23  & $\Delta R( \nu^{max},$ $jet 1)$ \\ 
    6 & $MetSig$             & 12 & $\Delta\phi( jet 2,$ \met$)$ &  18 & $\Delta R( lep,$ $jet 1)$ &  24 & $\Delta R( \nu^{max},$ $jet 1)$ \\ 
\bottomrule
  \end{tabular}
}
  \caption[Possible SVM Input Variables]{All the possible input variables used for the SVM training and optimization. See also Sections~\ref{sec:training_set} and~\ref{sec:var_desc} for a detailed description.}\label{tab:input_vars}  
\end{table}

\subsection{Optimization and Variable Selection}\label{sec:var_sel}

There is a wide literature about variable selection (see for example Ref.~\cite{Chen_Lin_2006} for a review of some methods). In Ref.~\cite{Nguyen-DelaTorre-PR10} the variable set is optimized during the training while, in Ref.~\cite{Mladenic2004}, multiplicative weights, ranging from 0 to 1 and tuned between successive training cycles, are associated to each input variable. In our case, as we use a figure of merit based on a statistical test and performed on the trained classifier, we were not able to use a feature selection system integrated into the training. We solved the problem by exploring the space of possible variables: we started from a reduced subset of all the possible inputs and then we increase it adding more variables to the most performing sets, as also done in Ref.~\cite{Avidan_feature_sel}.

Unluckily the brute-force search over all the possible combinations of variables across all the $C, \gamma$ phase space of a given SVM training is computationally unfeasible. 
Just the search over all the possible combinations of twenty-four variables requires a total of $16777215$ different SVM configurations.

\begin{figure}[!h]
  \begin{center}    
    \includegraphics[width=1.0\textwidth]{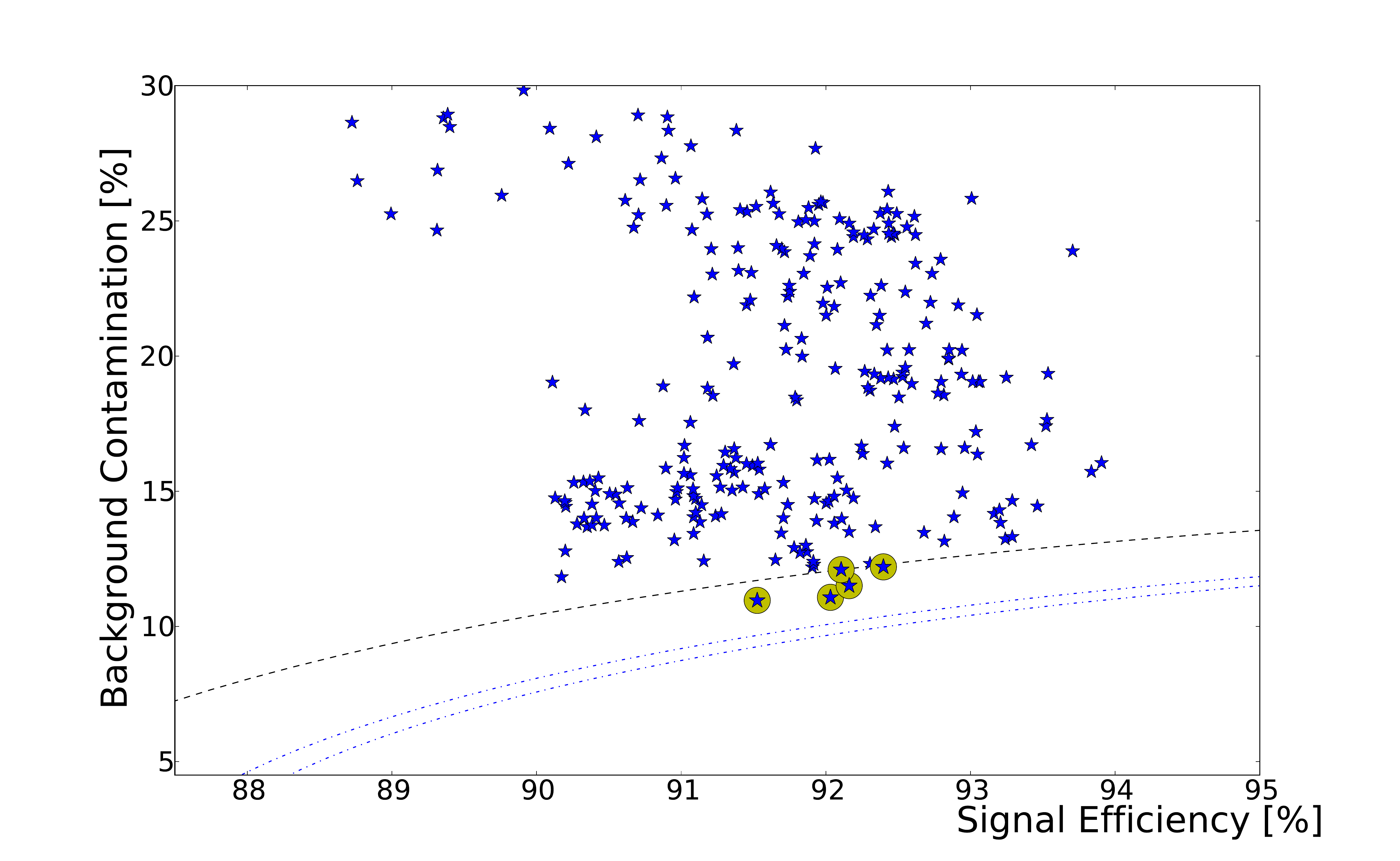}
    \caption[Signal Efficiency {\em vs} Background Contamination for Multiple SVM]{Performances of different SVM configurations, obtained by the combination of tree (out of twenty-four) input variables in the central region training, are displayed as blue stars on a signal-efficiency {\em vs} background-contamination scatter plot. The signal efficiency is derived from MC simulation while the background contamination is obtained from the two-component template fit of the SVM distance, $D$, described in Section~\ref{sec:method}. The five configurations closest to ideal performances according to Eq.~\ref{ideal_perf} are circled in yellow and selected for further iterations of the training with an increased number of input variables. We performed three iterations of the training, with three, six and eight input variables and the three dotted lines represent the distance from ideal performancesenclosing the five best configurations in each case. After the third iteration no more sensible improvement occurs.}\label{fig:discriminants}
  \end{center}
\end{figure}

\begin{figure}
	\centering
		\includegraphics[width=1.00\columnwidth]{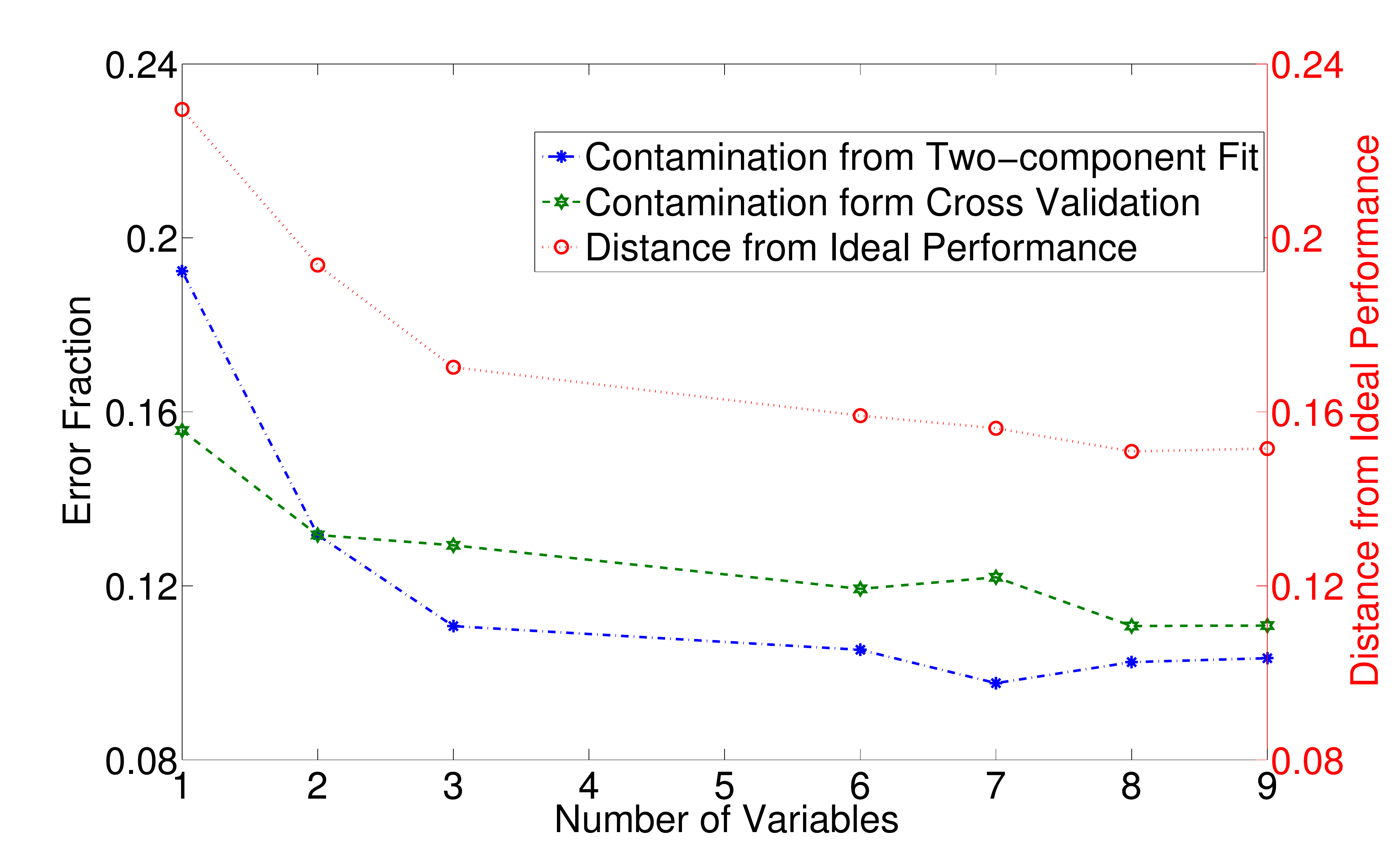}
	        \caption{Improvement of performance increasing the number of variables (for the central electron sample) as a function of three significant quantities: 
the fraction of mis-classified background events, or error fraction, as returned both from the $k$-fold cross validation ($f^{k-fold}_{Bkg}$) and from the two-component fit ($f^{Fit}_{Bkg}$), and the distance from the ideal performance (described by Eq.~\ref{ideal_perf}). Notice how all the curves of performance flatten within $1\div 2$\% when the number of variables approaches eight.}
	\label{improvement}
\end{figure}

To scan the most relevant sectors of the phase space we applied a factorized and incremental optimization:
\begin{itemize}
\item for all the configurations of three variables, we evaluate a grid of  $C, \gamma$ values in the intervals:
  \begin{equation}
   \log_2 C\in [-2,9] \quad\textrm{and}\quad \log_2\gamma\in[-4,7],
  \end{equation}
  where the use of a logarithmic scale allows to scan the parameters across several orders of magnitude. For each variable configuration we select only the best SVM training configuration, according to the result of the confusion matrix.
\item Then, for each {\em best} SVM and variable configuration, we perform a two-component template fit of the background and signal normalizations over the SVM variable $D$. We evaluate the  $\chi^2$ of the fit, reduced by the Number of Degrees of Freedom ($NDoF$), and we compare the fitted fraction of mis-classified background events, $f_{Bkg}^{Fit}$, against the one obtained from the $k$-fold cross-validation, $f_{Bkg}^{k-fold}$. The SVM under exam is rejected if:
  \begin{equation}
    \frac{\chi^2}{NDoF}>3\quad  \textrm{or} \quad\frac{f_{Bkg}^{Fit}}{f_{Bkg}^{k-fold}}>2\textrm{.}
  \end{equation}
  Notice that the quality of the fit is not directly optimized by the SVM training, so we are performing a consistency check of our classifier in an unbiased sample (data) with an independent technique.
\item  After the first iteration of the template fit cross check, we display all the remaining SVMs on a signal-efficiency ($\varepsilon_{Sig}$, derived from MC simulation) {\em vs} background-contamination ($f_{Bkg}^{Fit}$) scatter plot like the one in Figure~\ref{fig:discriminants}. The five best variable combinations according to the minimal distance, $d$, from ideal performance are selected for further processing, we add all the possible combinations of $1$, $2$ and $3$ variables to them for a second iteration. The distance $d$ is defined as:
  \begin{equation}\label{ideal_perf}
    d=\sqrt{\big(\varepsilon_{Sig}-\varepsilon^{Ideal}_{Sig}\big)^2 + \big(f^{Fit}_{Bkg}-f^{Ideal}_{Bkg}\big)^2}\textrm{,}
  \end{equation}
  with $\varepsilon_{Sig}^{Ideal}=1$, $f^{Ideal}_{Bkg}=0$.
\end{itemize}
After few iterations the {\it best result} does not improve significantly. In Figure~\ref{improvement} the performances are shown (for the central electron sample) as a function of three significant quantities: 
the fraction of mis-classified background events as returned both from the $k$-fold cross validation ($f^{k-fold}_{Bkg}$) and from the two-component fit ($f^{Fit}_{Bkg}$), and the distance from the ideal performance (described by Eq.~\ref{ideal_perf}). Notice how all the curves of performance flatten within $1\div 2$\% when the number of variables approaches eight.

The total number of SVM optimizations to be performed is the product of the number of explored $C$, the number of explored $\gamma$, and the number of combinations of variables. This equals approximately to $300000$, in the case of the three training steps explained before. The SVM optimizations have been divided in equal numbers between 121 CPUs and performed on a distributed analysis grid system~\cite{Sfiligoi2007, Bartsch04testingthe}. For the first training step ($3$ variables) the computation took 6 hours, 12 minutes and 22 seconds.

\subsection{Final SVM Configuration}\label{sec:svm_res}
The two optimal SVMs obtained from the central (superscript $c$) and forward (superscript $f$) electron training sets are defined by the $C$ and $\gamma$ hyper-parameters and by a combination of input variables. 
In particular the values of the hyper-parameters are:
\begin{equation}
C^c = 7, \: \gamma^c = -1 \quad\textrm{ and } \quad
C^f = 8, \: \gamma^f = -1.
\end{equation}
The specific input variables giving the best performances are reported in Table~\ref{tab:svm_fin_vars}, eight for central-electron selection and six for the forward electron selection. 
\begin{table}
  \begin{center}   
  \begin{tabular}{lccc}
    \toprule
    \multicolumn{4}{c}{ Final SVM Input Variables}\\
    \midrule
    Central SVM: &  $M^W_T$ & \metr &  ${\not}{p_T}$\\ 
                 & $MetSig$ & $\Delta \phi({\not}{p_T},$ \met) & $\Delta \phi(lep,$ \met$)$ \\
                 & $\Delta R(\nu^{Min},$ $lep)$  & $\Delta \phi(jet 1,$ \met)  & \\
    \midrule
    Forward SVM: & $M^W_T$ & \metr & ${\not}{p_T}$ \\
                 &  $MetSig$ &  $\Delta \phi({\not}{p_T},$ \met) & $\Delta \phi({\not}{p_T},$ \metr)  \\
    \bottomrule
  \end{tabular}
  \caption[Final SVM Input Variables]{Final input variables used for the configuration of the {\em central} and {\em forward} SVM multi-jet discriminants.}\label{tab:svm_fin_vars}
  \end{center}
\end{table}

Although the two training sets are selected independently and they present different kinematics, a certain degree of similarity arises in the final configurations as five of the final input variables
are in common between the two SVMs.  One of them, the $M^W_T$ is closely related to the kinematic of the $W$ decay, others, like the \metr ~and the ${\not}{p_T}$, give independent information about the reconstructed neutrino. With the $MetSig$ and the $\Delta \phi({\not}{p_T},$ \met), the angular information between the reconstructed quantities is exploited. 

The order of appearence of the variables during the factorized and incremental optimization may also give a qualitative information about the signal to backgrond discriminative power of them. For example the $M^W_T$  variable was always present in all the configurations coming from the first optimization cycle.  The $MetSig$, the \metr ~and the ${\not}{p_T}$ variables appear also to be relevant as they were often present in the first and second optimization cycles. 

As none of the input variables relies on the specific electron identification algorithm definition, the same SVM classifier can be applied in channels where the $W$ decay is selected by different lepton reconstruction criteria. An example of this usage is reported in Ref.~\cite{fede_sforza_tesi}: in this case the central SVM discriminant has been used in the multi-jet rejection for channels with muons in the final state and with leptons identified using only the tracker of the CDF detector.

\begin{figure}[!h]
  \begin{center}    
    \includegraphics[width=1.0\textwidth]{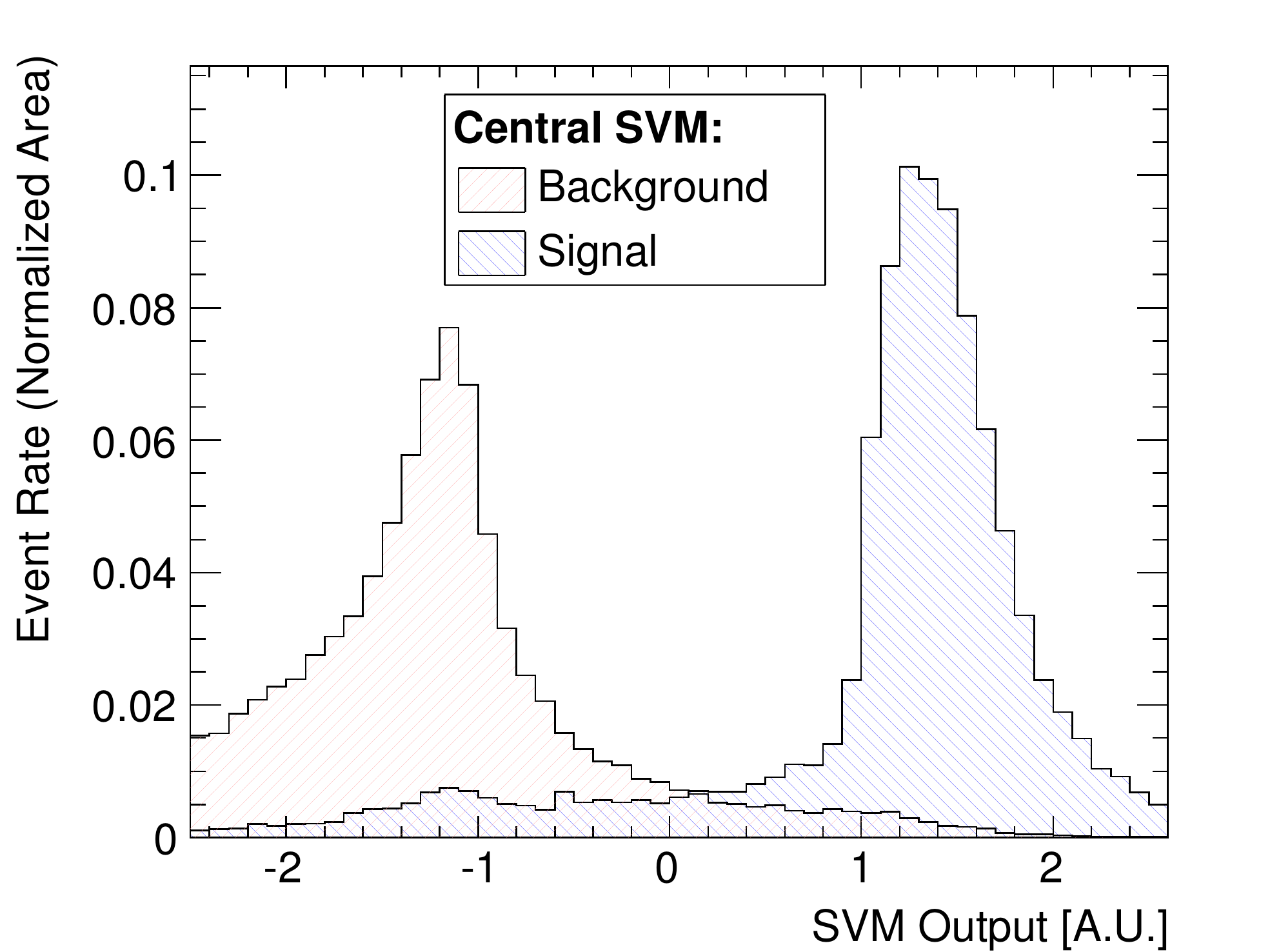}
  \caption[Final SVM Discriminants for Central Detector Region]{Distribution of the SVM variable $D$, described in Eq.~\ref{eq:distance}, for the central SVM discriminant obtained from the optimization process. Multi-jet background model is shown in red, $W+$jets MC signal is shown in blue.}\label{fig:finalcentral_svm}
  \end{center}
\end{figure}

\begin{figure}[!h]
  \begin{center}    
    \includegraphics[width=1.0\textwidth]{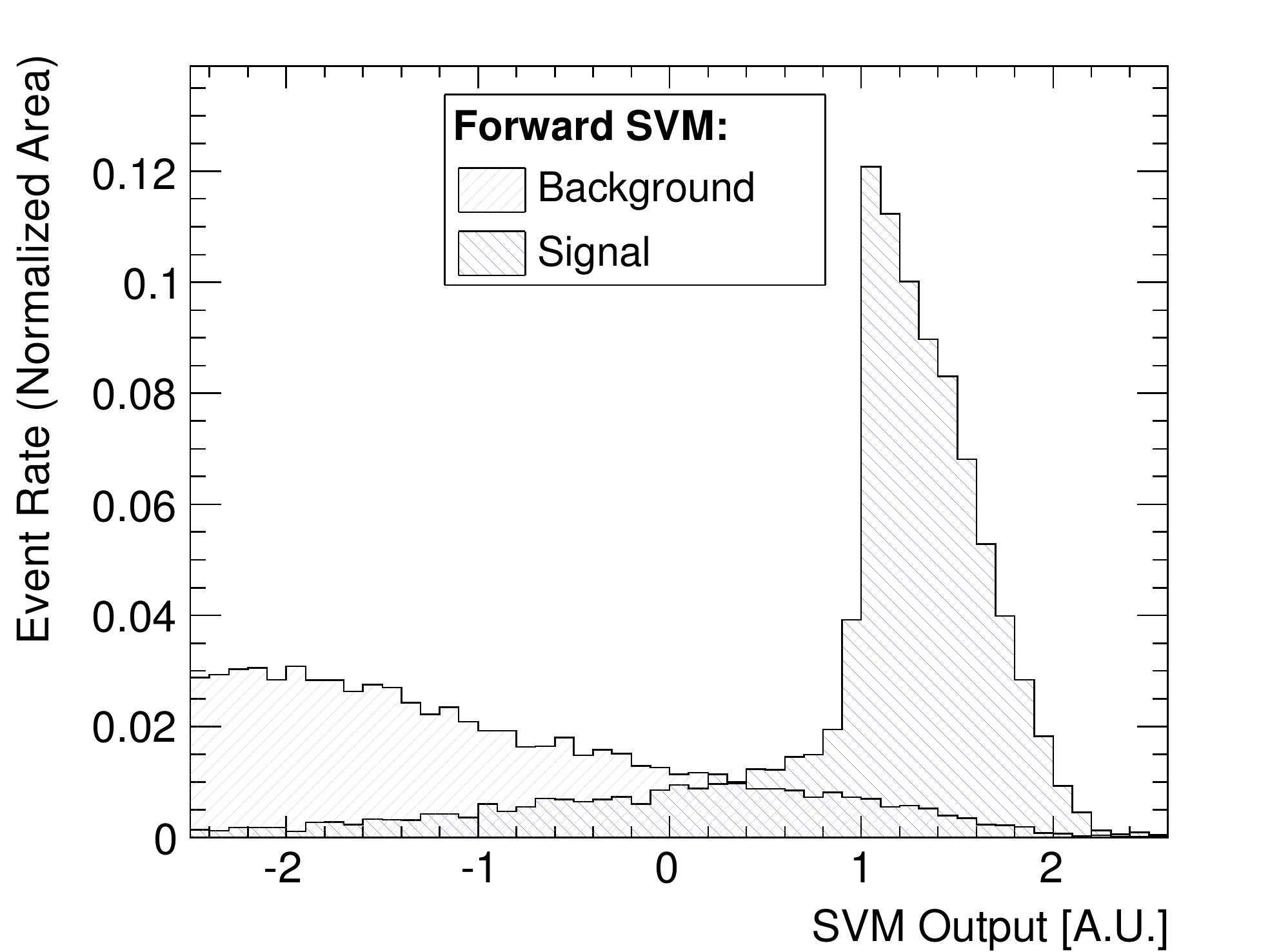}
  \caption[Final SVM Discriminants for Forward Detector Region]{Distribution of the SVM variable $D$, described in Eq.~\ref{eq:distance} for the forward SVM discriminants obtained from the optimization process. Multi-jet background model is shown in red, $W+$jets MC signal is shown in blue.}\label{fig:finalforward_svm}
  \end{center}
\end{figure}

Another advantage of the approach presented in this paper is the possibility to exploit the full shape information of the SVM output variable $D$. Figures~\ref{fig:finalcentral_svm} and~\ref{fig:finalforward_svm} show the shape of the central and forward SVM discriminants for the multi-jet background models and the $W+$jets signal. A lower background contamination can be achieved increasing the SVM selection threshold from the zero value. 

Table~\ref{tab:svm_fin_perf} reports the evaluation of the performance as given by the two elements of the confusion matrix useful in a multi-jet background rejection problem: the signal efficiency (obtained from the training set) and the background contamination (obtained from the template fit procedure) both evaluated for a SVM selection threshold of $D = 0$, used during the training optimization, and of $D = 1$, as an additional example. The different performance of the central and of the forward SVM discriminants arise from the different shape of the background distributions (as shown in Figures~\ref{fig:finalcentral_svm} and~\ref{fig:finalforward_svm}). This is due to the specific kinematic of the multi-jet background in the forward region, characterized by highly boosed objects reconstructed in a less finely segmented detector area.

The performance of our multi-jet rejection algorithm can also be compared to other methods used in similar contexts. The $WH\to e\nu$ plus two jets associate production and decay (with a Higgs boson mass of $115$~GeV$/c^2$) can be used as a common signal reference process to evaluate the effective performance of the algorithms.
A first direct comparison is possible within the CDF collaboration against a previous SVM based method~\cite{CHEP_svm} used in several analyses~\cite{wh75_PRD, wh94_PRL, cdfDiblvHF_Mjj2011, cdf_zprime2012}. The new multi-jet rejection strategy improves the $WH$ signal acceptance by $5$\% allowing the same multi-jet background contamination. Then it is possible to compare results presented in this paper to the multivariate strategy applied by the D0 collaboration in the same decay channel. Ref.~\cite{d0_WH_8.5} describes a multi-jet rejection power of $75$\% for a very loose operating point of the algorithm which allows a $WH$ signal efficiency of $97$\%. For the same signal efficiency, the approach described in this paper rejects approximately $80$\% of the multi-jet background. The last comparison is done against the recent results of the Atlas~\cite{atlas_VH2012_conf} and CMS~\cite{cms_VH_2012_conf} collaborations. In these case the running conditions (for example the instantaneous luminosity and the pile-up of the events) are more challenging,  therefore the achievement of a comparable multi-jet background contamination is an extremely good result. However a lower signal efficiency is expected from the use of tight selection criteria: for example the presence of large \met ~or large $M^W_T$ in the event.

It is possible to conclude that, at today, the presented result is superior to any other multi-jet rejection strategy applied for the same final state (i.e. $W\to e\nu$ plus two jets) at hadron collider experiments.
\begin{table}
  \begin{center}   
{
\renewcommand{\arraystretch}{1.2}
  \begin{tabular}{lcccc}
    \toprule
     &  \multicolumn{2}{c}{Signal Efficiency} & \multicolumn{2}{c}{Background Contamination} \\
    & $D\ge 0$ & $D\ge 1$ & $D\ge 0$ & $D\ge 1$\\
    \midrule
    Central SVM: & $0.935 \pm 0.002$ & $0.837 \pm 0.002$& $0.101\pm 0.004$ & $0.043\pm 0.002$\\
    Forward SVM: & $0.908 \pm 0.003$ & $0.753 \pm 0.004$& $0.302\pm 0.004$ & $0.135\pm 0.002$\\
    \bottomrule
  \end{tabular}
}
  \caption[Final SVM Performance]{Performance of the {\em central} and {\em forward} SVMs final configurations evaluated for two selection thresholds: $D = 0$ (the value used in the training optimization) and $D =1$. The signal efficiency is derived the MC selection while the background contamination is derived from the two-component template fit procedure described in Section~\ref{sec:method}. Errors are statistical only.}\label{tab:svm_fin_perf}
  \end{center}
\end{table}

\subsection{Example of Use for Background Estimate}

An example of the use of the developed SVM discriminant is reported in Ref.~\cite{fede_sforza_tesi}. In particular both the possibility to move the SVM selection threshold and the fit of the multi-jet contamination over SVM output variable $D$ are used. 

\begin{figure}[!ht]
\begin{center}
\includegraphics[width=1.0\textwidth]{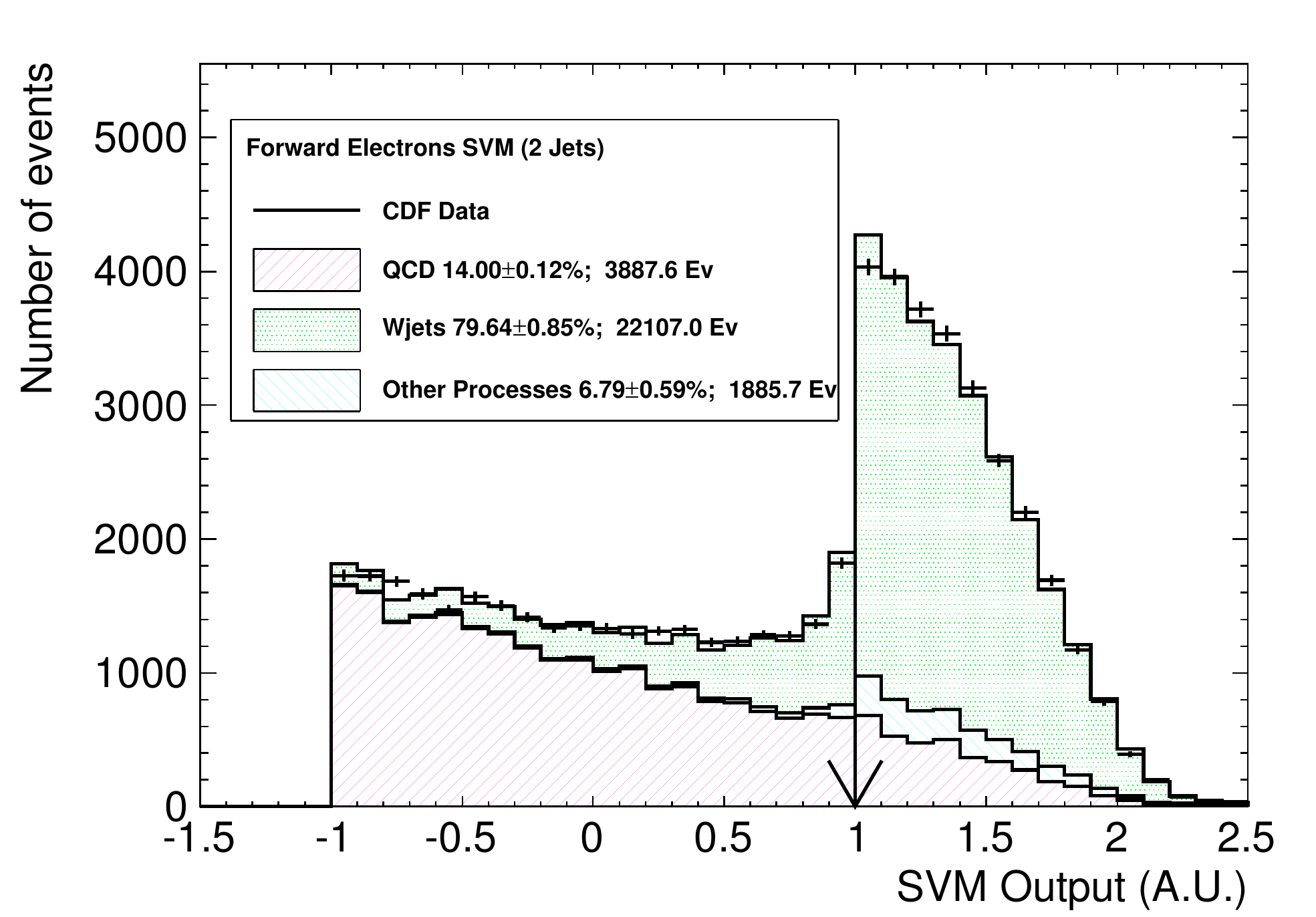} 
\caption{Contribution of the different physics processes to the shape of the SVM output distribution $D$ used during the {\em forward} electron sample selection. The multi-jet, or QCD, background fraction (in magenta) is extracted from the fit together with the total $W$ plus jets component (in green). The remaining physics processes (in light blue) are normalized to the expected production cross sections and acceptances. The SVM selection threshold for the final signal region identification is $D = 1$ instead that the standard value of $D = 0$. This further decreases the QCD contamination in the final analysis sample.}\label{fig:svm_phx_fit}
\end{center}
\end{figure}

Figure~\ref{fig:svm_phx_fit} shows how the different physics processes contribute to the shape of the SVM discriminant used in the selection of the forward electron sample. The multi-jet, or QCD, fraction is extracted from the fit together with the total $W$ plus jets component. The SVM threshold used for the final signal region identification is $D = 1$ instead of the standard value of $D = 0$. This further decreases the multi-jet contamination and reduces the impact of the systematics related to the estimate of such background.

\section{Conclusion}

In this paper we present an innovative method for training optimization and variable selection for a SVM multivariate classifier in the problematic case of biased and statistically limited training samples.

The optimization of the classifier is possible thanks to the feedback of a signal-background template fit performed on a validation sample of unclassified events over the SVM output distribution $D$ (defined in Eq.~\ref{eq:distance}). By construction, the SVM output $D$ is a variable sensitive to the signal and background contamination.

The optimization algorithm was then applied to an actual hadron collider physics case. The problem of multi-jet rejection in the $W$ plus jets channel has been analyzed in two different cases: a $W$ boson decaying to an electron identified in the central and in the forward region of the CDF detector and selected together with two or more hadronic jets. 

The optimization algorithm allowed to scan a pool of twenty-four input variables to obtain the minimal combination of them giving the lower background contamination and the higher signal efficiency. The resulting classifiers exploit eight and six input variables combinations, respectively for the {\em central} and {\em forward} SVM classifiers, not directly related to the lepton identification algorithm. The performance is superior to any present multi-jet rejection algorithm applied to the same final state. 

The CDF II dataset and the specific multi-jet rejection analysis have been a perfect test-bench for the SVM optimization algorithm. However the procedure can be exported to any other problem where the use of multivariate techniques is required but no accurate simulation is available or possible.

\section*{Acknowledgments}

We would like to thank the University Research Association for the support given to F. Sforza with the 
Visiting Scholars Program at FNAL in Spring 2010, Dr. Giorgio Chiarelli and Dr. Sandra Leone for the support, the guidance and the discussions that contributed to this paper, all the CDF Higgs Discovery Group for the feedback and the suggestions that improved our research and, in particular, Dr. H. Wolfe for the review of the manuscript.


\bibliographystyle{elsarticle-num-names.bst}

\end{document}